\documentclass[preprint,12pt]{elsarticle}
\usepackage{graphicx}
\usepackage{natbib}
\usepackage{txfonts}
\usepackage[dvips]{color}
\usepackage{upgreek,latexsym}

\journal{Icarus}

\begin{document} 

\begin{frontmatter}
\title{Scheila's Scar:\\ Direct Evidence of Impact Surface Alteration on a Primitive Asteroid} 
\author[UMD]{D. Bodewits}
\author[MPS]{J.-B. Vincent}
\author[UMD]{M.\ S.\ P. Kelley}


\address[UMD]{Department of Astronomy, University of Maryland, College Park MD 20742-2421, USA, dennis@astro.umd.edu, msk@astro.umd.edu}
\address[MPS]{Max-Planck-Institut f\"{u}r Sonnensystemforschung, Max-Planck-Str. 2, 37191 Katlenburg-Lindau, Germany, vincent@mps.mpg.de}

\begin{abstract}
Asteroid (596) Scheila was the first object for which the immediate aftermath of an inter-asteroidal collision was observed. In Dec. 2010, the 113 km-sized asteroid was impacted by a smaller asteroid of less than 100~m in diameter. The scale of the impactor was established by observations of fading ejecta plumes. Comparison of the light curves obtained before and after the impact allowed us to assess how much of Scheila's surface was altered. Cratering physics based on the impactor size suggests that the size of the affected area is larger than expected, (effective radii of 3.5 -- 10 km depending on the change in the albedo of the surface). Similar but more localized albedo changes have been observed on Vesta and the Martian moons, but are not understood. Empirical laws describing ejecta blankets however indicate that at distances between 3.5 -- 10 km from the crater, Scheila's surface would be covered by a thin layer 2 mm to 2 cm thick. This dusting, possibly mixed with bright impactor material may be enough to explain to observed brightness increase. Our results show that sub-critical impacts may play a significant role in processing the surfaces of asteroids. The large effect of small impacts on asteroidal light curves complicate shape modeling.
 
\end{abstract}


\end{frontmatter}
%

\section{Introduction}

\noindent Early December 2010, an unexpected dust cloud was discovered around the asteroid (596) Scheila \citep{2010IAUC.9188....1L}.  The asteroid's brightness had increased by almost 1 magnitude, and archival Catalina Sky Survey observations showed that the activity was triggered between 2010 Nov. 11 and 2010 Dec. 11 UT \citep{2010IAUC.9188....1L}. Quick follow-up observations  and spectroscopic investigations found no evidence for OH or other gas in near-UV \citep{2011ApJ...733L...3B}, optical \citep{2012ApJ...744....9H}, and radio wavelengths \citep{Howell:2011p2648}. IR investigations excluded the presence of surface ice on Scheila \citep{2011ApJ...737L..39Y}.  \citet{2011ApJ...733L...3B} reported the near-UV/optical color of the surface was red, and the dust was neutral, while  \citet{2012ApJ...744....9H} reported that the color of the dust was consistent with the red surface at longer wavelengths. Color differences between dust and surface are not unexpected, and a blued UV-optical dust color is consistent with the difference between scattering from a surface, and scattering from an optically thin collection of small dust grains \citep{2011ApJ...733L...3B}. The dust plumes rapidly faded within the next months \citep{Jewitt:2011p2737} and the last reported observations of a thin trail that likely consists of large dust grains were achieved on March 2, 2011, using the Subaru telescope \citep{Ishiguro:2011to}. 

These observations indicate (596) Scheila, a large asteroid of 113~km in diameter, was impacted by a smaller asteroid. The observed morphology suggested an oblique impact  where a large plume was associated with downrange ejecta and a limb-brightened cone with material ejected vertically due to the mechanical excavation by shockwaves generated by the impact \citep{2011ApJ...733L...3B,Ishiguro:2011uh,2011ApJ...738..130M}. The observed fading of the plumes is consistent with ejecta moving away from the asteroid and then being pushed along the anti-solar direction by solar radiation pressure. Depending on assumptions on density and grain size distribution, it was estimated that the dust plumes contained 0.3 -- 6 $\times 10^8$ kg of dust \citep{2011ApJ...733L...3B,2012ApJ...744....9H,Ishiguro:2011to,Jewitt:2011p2737}. Assuming that this dust was ejected at velocities larger than 68~m$\cdot$~s$^{-1}$ to overcome Scheila's gravity, and that mass of the plume corresponded to 1 -- 10$\%$ of the total excavated mass, all studies agreed that the outburst was caused by an impactor of about 30 -- 80 meters, and likely resulted in a crater with a diameter between 300 -- 800 m. Reverse modeling suggested this impact occurred around Nov. 27 -- Dec. 3, 2010 \citep{2011ApJ...738..130M,Ishiguro:2011uh}. Scheila thus provided the first non-ambiguous detection of the immediate aftermath of an inter-asteroidal collision and as such presents a unique experiment, since both a direct measure of the scale of the impact is available (through the observed dust plumes) and because the impact was not catastrophic, so that the effect of the impact on Scheila can be studied.

   \begin{figure*}
   \hfill
   \begin{minipage}[t]{0.48\textwidth}
   \centering
   \includegraphics[width=6.5cm]{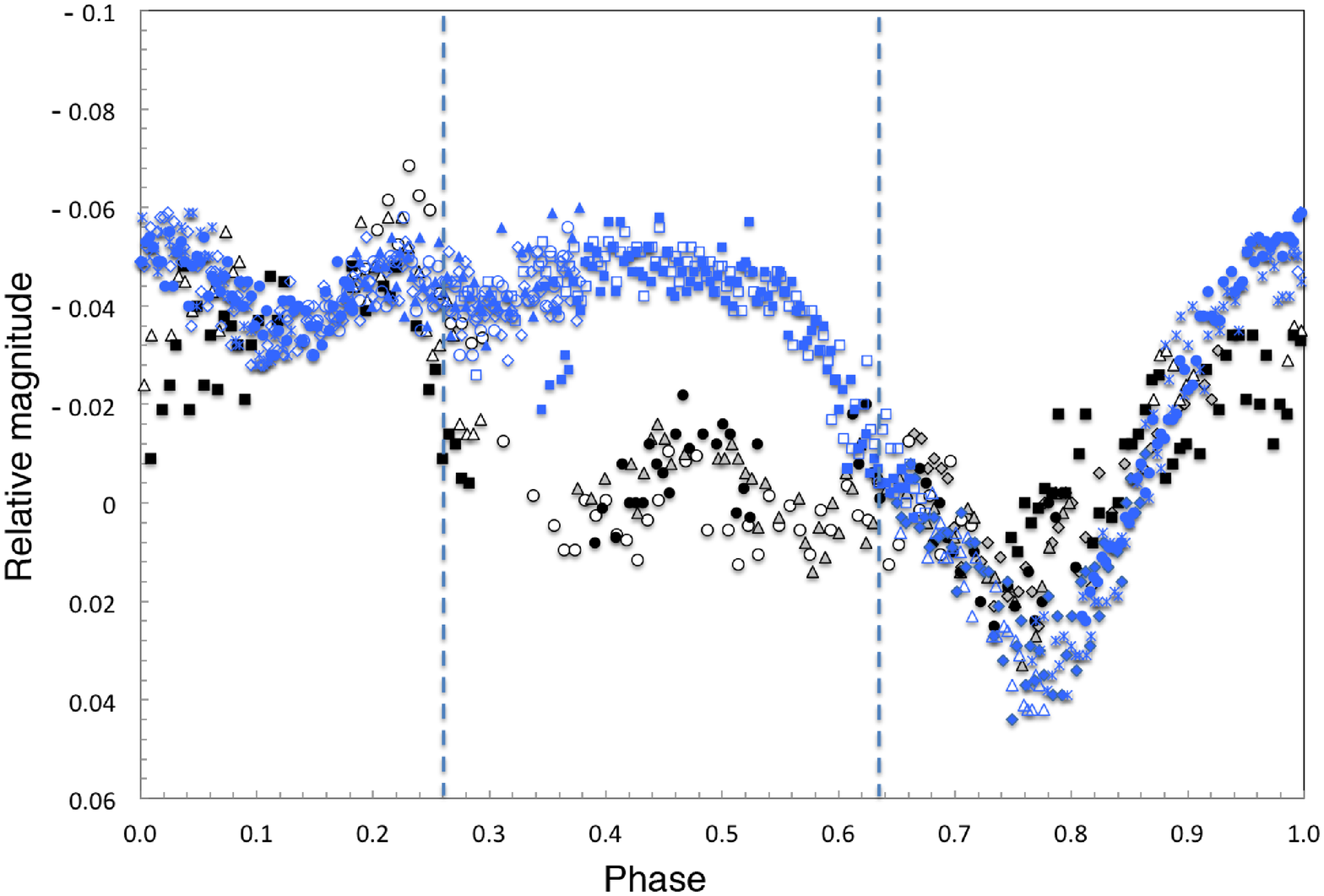}
   \caption{The light curve of Asteroid (596) Scheila before (black and white, \citet{Warner:vz}) and after the impact (blue symbols, \citet{Ishiguro:2011to}). Different symbols indicate different observing days. Before impact: \textcolor[gray]{0.7}{$\blacklozenge$} - 1/10/2006; $\circ$ - 1/13/2006; \textcolor[gray]{0.7}{$\blacktriangle$} - 1/21/2006; $\vartriangle$ - 1/22/2006; $\bullet$ - 1/23/2006; $\blacksquare$ - 1/24/2006. Post impact: \textcolor{blue}{$\blacklozenge$} - 12/17/2010; \textcolor{blue}{$\circ$} - 12/20/2010; \textcolor{blue}{$\vartriangle$} - 12/23/2010; \textcolor{blue}{$\blacktriangle$} - 12/26/2010; \textcolor{blue}{$\ast$} - 1/2/2011; \textcolor{blue}{$\square$} - 1/3/2011; \textcolor{blue}{$\blacksquare$} - 1/7/2011; \textcolor{blue}{$\bullet$} - 1/8/2011; \textcolor{blue}{$\lozenge$} - 1/28/2011. The dashed vertical lines indicate the part of the lightcurve that was affected most by the impact.
   }
              \label{Fig_LC}%

   \end{minipage}
   \hfill
   \begin{minipage}[t]{0.48\textwidth}
   \includegraphics[width=6.5cm]{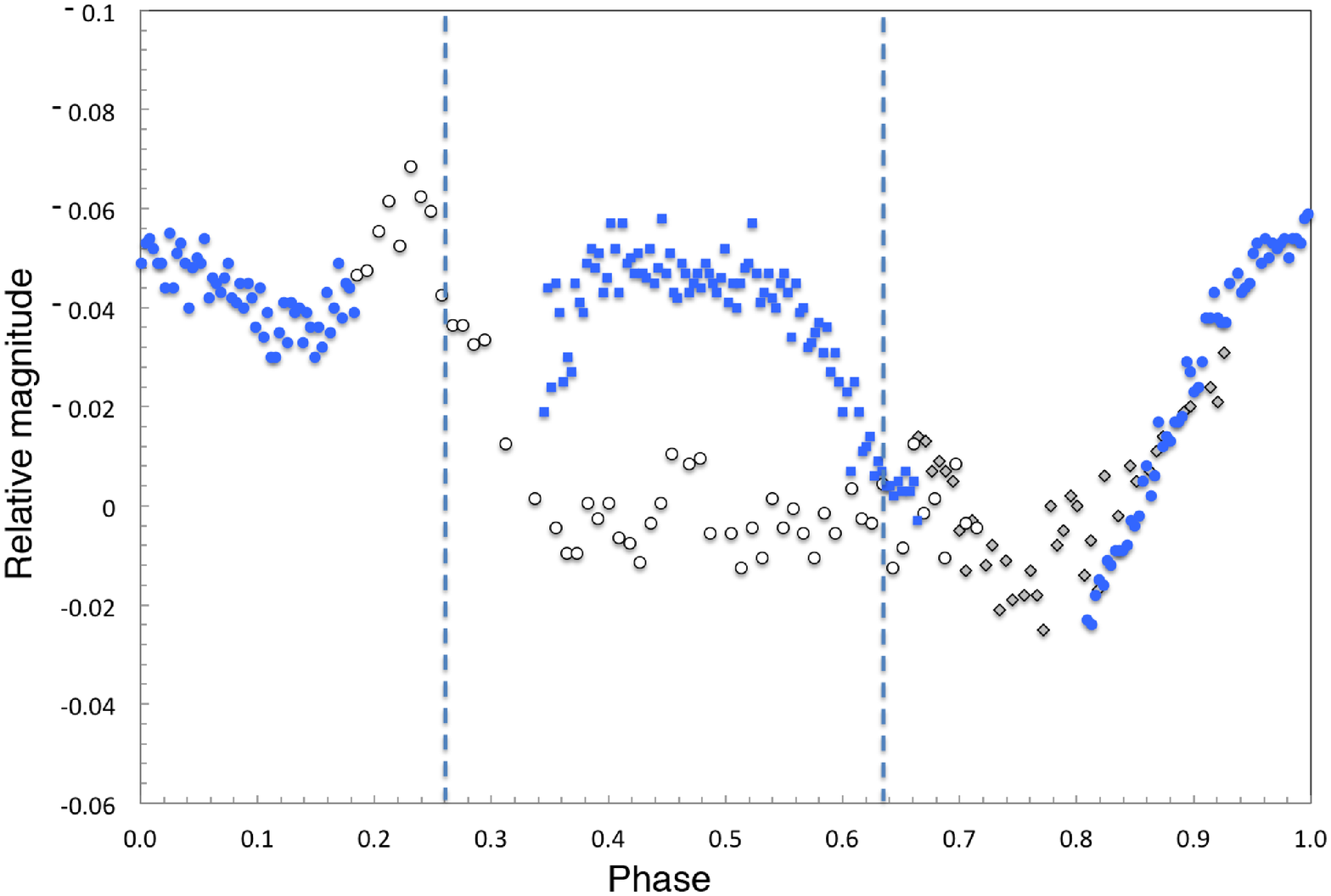}
   \caption{Subset of the light curve measurements of Asteroid (596) Scheila measured at phase angles of approximately 11 degrees. Symbols are equal to those used in Fig. 1;  Different symbols indicate different observing days. Before impact: \textcolor[gray]{0.7}{$\blacklozenge$} - $\phi$ = 10.7 deg; $\circ$ - $\phi$ = 9.97 deg; Post impact: \textcolor{blue}{$\blacksquare$} - $\phi$ = 11.26 deg; \textcolor{blue}{$\bullet$} - $\phi$ = 11.02 deg. The dashed vertical lines indicate the part of the lightcurve that was affected most by the impact.
   }
              \label{Fig2}%
              \end{minipage}
              \hfill
    \end{figure*}

Asteroid Scheila has a T-type spectrum \citep{2009Icar..202..160D} and orbits the Sun in the outer-main belt with a 5.0 year period, an eccentricity of 0.16, and a semimajor axis of 2.93 AU. As a primitive asteroid, Scheila probably has a surface composition that is close to carbonaceous meteorites \citep{2009Icar..202..160D, 2011epsc.conf.1109L, 2011ApJ...737L..39Y}. Since their formation, the geology and mineralogy of the surfaces of large asteroids are essentially dominated by impacts. Impacts excavate fresh material from the parent body interior, which is subsequently exposed to space weathering effects such as solar wind irradiation and micrometeoroid bombardment. The surface evolution of primitive asteroids is not well studied because only one has been visited by a space mission (253 Mathilde; \citealt{Veverka1999}), and few recent families are known, hampering a comparative study that might be further blurred by compositional differences amongst C-complex asteroids. As a result, the limited comparative studies in literature have apparently contradicting results \citep{2005Nesvorny,2006ApJ...647L.179L}.  However, the Martian satellites, Phobos and Deimos, both imaged in detail by spacecraft, have D-type spectra, and can be used to help infer the properties of the outer-main belt D-types. The aim of this paper is to explore the effect of a relatively small impact on the optical surface properties of Asteroid (596) Scheila. The main evidence of this effect comes from photometric measurements before and after the impact (Sec. 2). We will next review the size of the crater and the area of the resulting ejecta blanket based on recent asteroid missions (Sec.~3). We will then discuss the geological implications which we hope will help to improve observational studies and theoretical models alike (Sec.~4).


\section{Lightcurve}

\noindent Measurements close to opposition in 2006 were obtained by B.~Warner using a combination of 0.35~--~0.5-m telescopes at Palmer Divide Observatory \citep{Warner:vz,2011MPBu...38...96W}. Post-impact photometry were acquired between Dec. 2010 and Jan. 2011 by H.~Hamanowa with the Hamanowa Astronomical Observatory  \citep{Ishiguro:2011to} and by \citet{Husarik:2012wc} at the Skalnate Pleso Observatory. The photometry data sets consist of several nights of observations to acquire complete coverage of Scheila's 15.9-h rotation period.  Using the data by \citet{Warner:vz,2011MPBu...38...96W} and \citet{Ishiguro:2011to} we constructed phase curves, scaling individual nights to a continuous light curve, which was then folded with a period of 15.877~h \citep{2011MPBu...38...96W}. The relative phase pre- and post-impact data were aligned by matching their respective minimum, which is a clear feature in both light curves. It is clear that the impact in 2010 significantly altered Scheila's light curve (Fig~\ref{Fig_LC}). Between a phase of $\phi = 0.26 - 0.63$ the lightcurve has increased by as much as 0.03 -- 0.06 magnitudes. The change is the largest around  $\phi = 0.36$. The lower estimate of 0.03 is based on the lightcurve of \citet{Husarik:2012wc}, which seems to suggest a smaller change than the \citet{Ishiguro:2011to} data. A detailed comparison of the two lightcurves, their discrepancies, and how each was constructed, is outside the scope of this paper.

To investigate whether the observed differences could be explained by differences between the observing geometries of the two observations, we compare observations acquired with comparable phase angles (around 11 degrees) in Fig.~\ref{Fig2}.  No relative scaling of the data subsets can make a consistent light curve; there will always be an offset between the two.  Note that given the measured phase darkening of the asteroid \citep[$G=0.08\pm0.06$ in the IAU $HG$ magnitude system;][]{2011MPBu...38...96W} and based on the ~1 degree difference in phase angles, a 0.04~--~0.05~mag offset may be expected.  However, the post-impact portion at 0.4 to 0.6 phases was observed at larger phase angles, and therefore should be darker than the corresponding pre-impact light curve.  Moreover, the post-impact change was observed over a wide range of phase angles (11~--~16 degrees, with the best coverage at 11 and 12 degrees).  The post-impact change is therefore most likely not attributable to a difference in the observing geometry. Considering the diameter of the impactor (30~--~80~m) compared to that of Scheila (113~km), one would not expect any change in the asteroid's shape or rotation period. \cite{Ishiguro:2011to} found that the post-impact rotation period agreed well with the one found by \citet{Warner:vz} five years earlier. We, along with \citet{Ishiguro:2011to}, therefore conclude that the change in lightcurve indicates that the impact has changed the optical properties of at least part of the surface. 

It is difficult to convert the change in lightcurve to a real surface area without knowing the impact latitude, the shape, and the spin axis orientation of Scheila, but we can make some reasonable assumptions. Because of the low phase angle of the observations, nearly the entire visible surface is illuminated. However, we do see clear periodic variations in the photometry. It is thus very unlikely that the spin axis of Scheila is aligned with our line of sight because we would then always see the same region of the surface at any time of the rotational phase. It is very reasonable to consider a spin axis pointing away from the observer, although we cannot constrain how much. 

\begin{figure*}
   \centering
   \hfill
   \begin{minipage}[t]{0.48\textwidth}
   \centering
   \includegraphics[width=7cm]{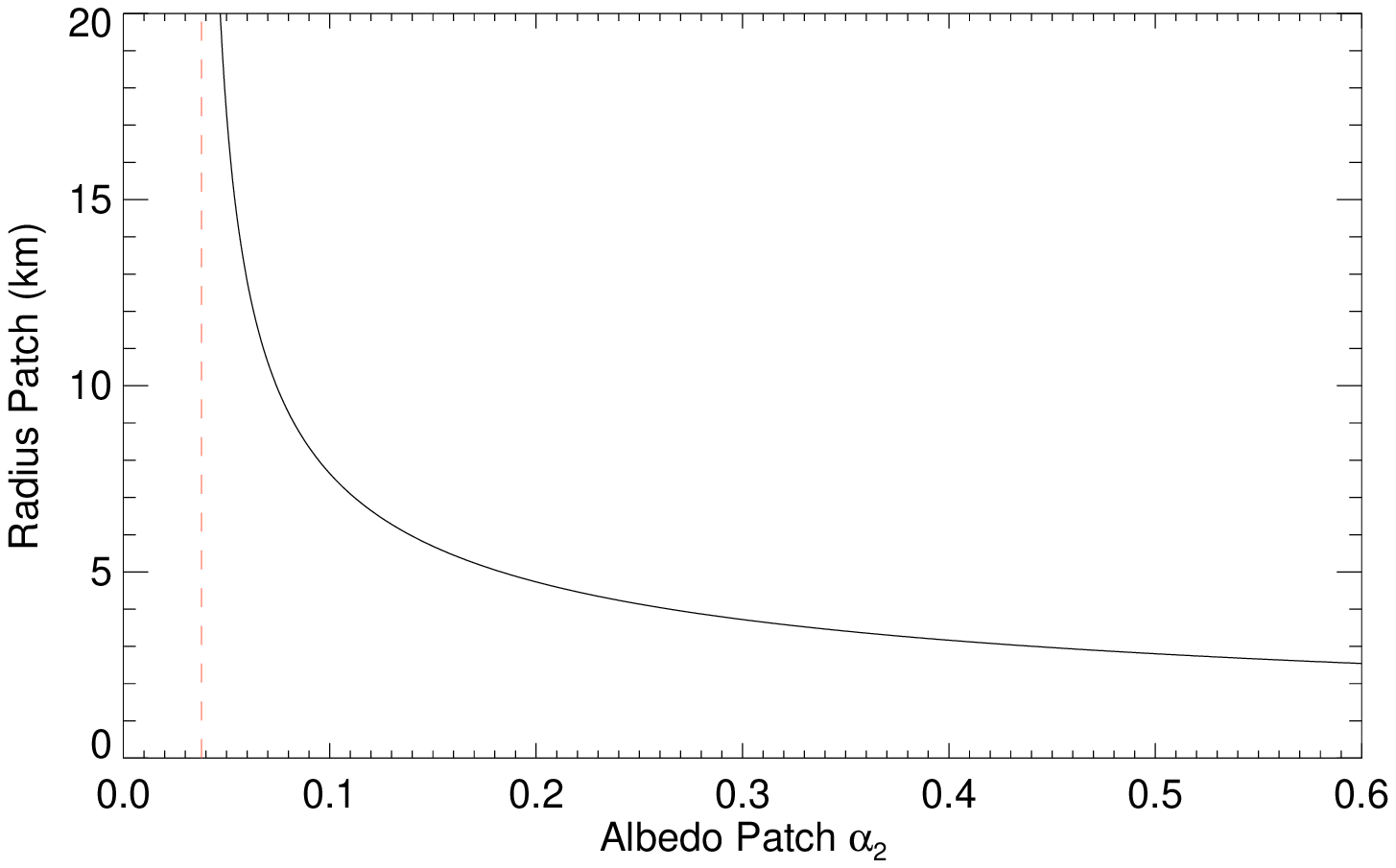}
   \caption{Relation between radius and geometric albedo of the bright patch on Scheila. The asteroid itself has an albedo of 0.038, indicated by the dashed red line.
   }
              \label{Fig3}%
   \end{minipage}
   \hfill
   \begin{minipage}[t]{0.48\textwidth}

   \centering
   \includegraphics[width=7cm]{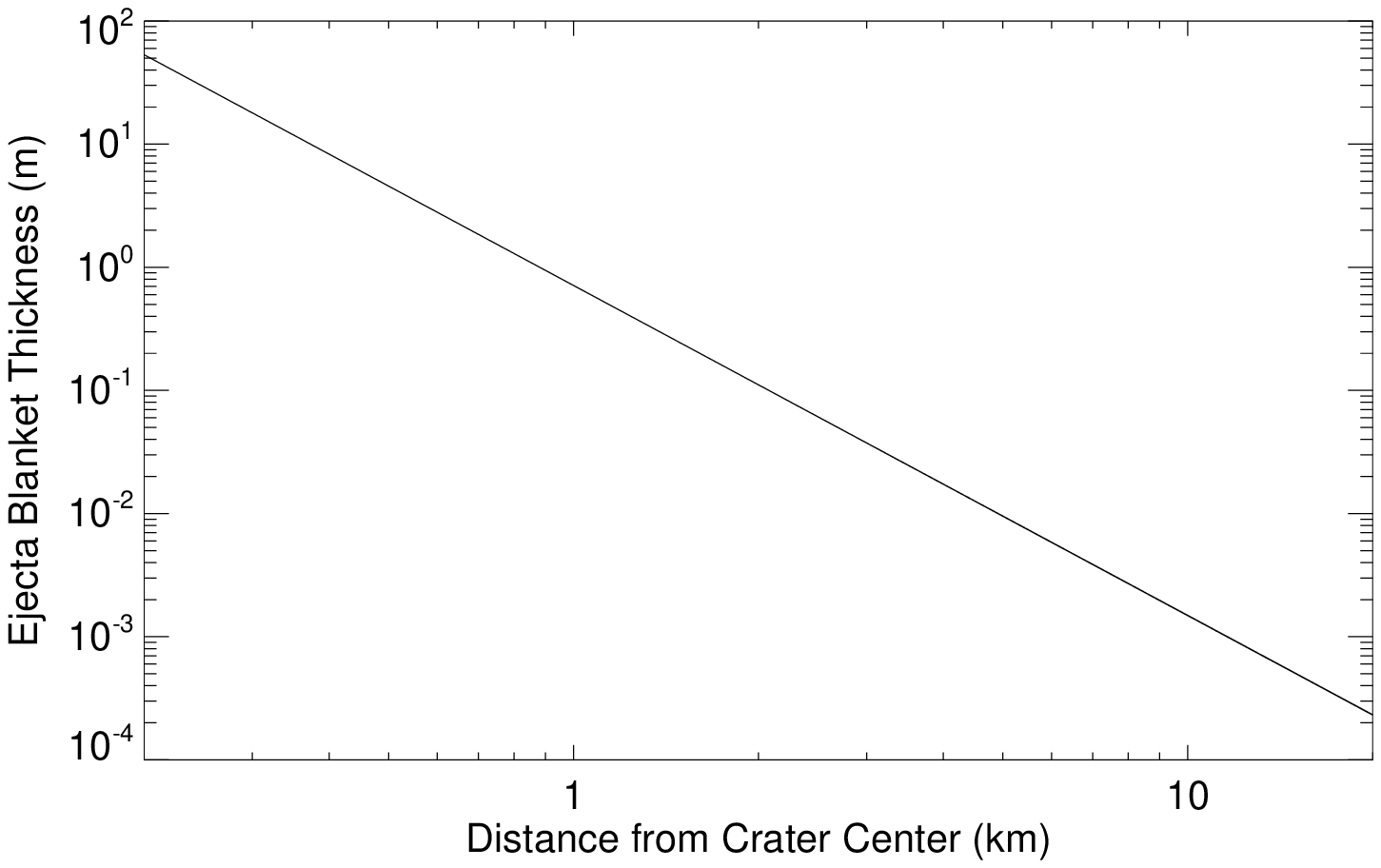}
   \caption{The predicted thickness of the ejecta blanket as a function of distance to the crater center, assuming a crater of 720~m in diameter.}
              \label{Fig4}%
    \end{minipage}
    \hfill
    \end{figure*}

Taking these arguments into account we can safely assume that the changes observed in the lightcurve are due to a bright region entering the illuminated area, and remaining visible for a third of the full rotation. Figure \ref{Fig_LC} shows that the post impact surface is 3 -- 6\% brighter than before. If we name $A_1$ the original surface area (albedo $\alpha_1$ = 0.038) and $A_2$ the area of the bright patch (albedo $\alpha_2$), we can write
\begin{equation}
(A_1-A_2)\alpha_1 + A_2\alpha_2 = 1.03 \times A_1\alpha_1
\end{equation}
which simplifies in
\begin{equation}
A_2 = \frac{0.03}{\alpha_2/\alpha_1 - 1} \times A_1
\end{equation}
(Throughout the paper we take albedo to mean geometric albedo, and we will use the lower limit of the observed change in the lightcurve.) This relation between albedo and the radius of the bright patch is illustrated in Fig.~\ref{Fig3}. How $\alpha_2$ compares with $\alpha_1$ depends on what process drove the brightening of the surface material and thus depends on the properties of both target and impactor. We believe the range of reasonable albedo increases to be between factors of 2 -- 10 for endogenic and exogenic processes, respectively (Sec. \ref{source}). If the albedo of the patch is 2 times higher than the background surface material, it will have a radius of 10~km. Similarly, if it were 10 times brighter the patch would be 3.5 km in radius. For the reasons stated above, we assumed here that the maximum peak of the light curve corresponds to the full ejecta blanket in the field of view, near the sub-observer point. The patch radius is a minimum value and could be larger depending on the geometry of the observations. 

\section{Discussion}
    
\noindent The observed change in the light curve and the observed plumes provide two independent measurements of the effect of the impact on Scheila. In this section, we will discuss whether they are consistent with each other. Based on estimates of the amount of mass excavated, we discuss the possible size of the crater, the distribution of ejecta around it, and how this could have affected the albedo of the surface.

\subsection{Crater size}
\noindent As reviewed in the Introduction, several authors have measured the amount of material present in the plume of ejecta observed shortly after the impact and measured a mass of $0.3-6 \times 10^8$ kg. Depending on many parameters such as the impact velocity and geometry, porosity, strength of the target, and the collision geometry, the dust cloud represented between 1\% and 10\% of the total amount of excavated material \citep{2002aste.conf..443H,2011ApJ...738..130M, Ishiguro:2011to}. The rest of the dust fell back onto Scheila's surface, covering a significant area with a thin layer of dust (Sec. 3.2). The total amount of material excavated by the impact is therefore probably in the order of a few $10^{10}$~kg. Scheila, being a T-type asteroid, has unknown compositional and surface properties but its surface density is likely to be of the same order as other large rocky bodies, i.e. $2500$~kg$\cdot$m$^{-3}$. From this density we obtain a \emph{maximum} total volume of excavated material in the range of $1.2-24 \times 10^6$~m$^3$.
This volume corresponds to the material lying above about 1/3$^{rd}$ of the transient crater depth \citep{melosh89} and can be calculated with the following integral:
\begin{equation}
V = \int_0^{d/3} \pi r^2 (1-\frac{z^2}{d^2}) dz
\end{equation}
Where $d$ is the depth of the crater and r its radius.
The depth of a transient crater is about 1/4$^{th}$ of its diameter, i.e. half of its radius. Hence we can simplify the volume formula and after integrating we obtain the crater diameter from the volume ejected:
\begin{equation}
D = 2 \times \sqrt[3]{\frac{81}{13}\frac{V}{\pi}}
\end{equation}

\noindent With this formula we obtain a transient crater diameter of 260 m to 720 m, likely resulting in a final crater of comparable size. 

\subsection{Ejecta blanket}
\noindent The ground-based observations show that 37\% of the rotational lightcurve has been modified by the impact (Fig.~1), likely corresponding to an area of 9 -- 28 km in diameter. From the brightness of the observed plumes we estimated the crater to be no larger than several hundreds of meters in diameter and ruled out direct detection of the crater or changes to Scheila's shape or rotation period based on the scale of the impact (Sec.~2). This suggests that the changes in the lightcurve are due to a change in the optical properties of Scheila's surface, most likely by an ejecta blanket. This raises several further questions: is the size of the bright spot in accordance with typical ejecta blankets, how thick is the ejecta blanket, and how thick should it be to have a noticeable effect on the optical properties of the surface?

Impacts do not necessarily lead to ejecta blankets. \citet{Housen2012} suggest that porous bodies ranging in size from roughly a few tens to a few hundred km might suppress  ejecta because much of it is trapped within the craters. The lack of ejecta blankets around craters on Asteroid (253) Mathilde and the Saturnian satellite Hyperion is due to their high porosities: 50\% for Mathilde \citep{Veverka1999}, and over $40\%$ for Hyperion \citep{Thomas2007}. In these and similarly porous bodies, the energy from the impact compacts the asteroid surface, with little left over to eject material \citep{Housen2012}. The porosity of Scheila is not known, but primitive asteroids are generally believed to be porous. However for Scheila direct observations of the ejecta plumes exist and the estimated size of the crater is an orders of magnitude below the threshold crater size of $\approx$10 km, suggesting that energy loss through compaction is not important in this event \citep{Housen2012}.

A typical crater displays a continuous ejecta blanket until a distance of 2 to 3 times the crater radius, and discontinuous ejecta (for instance rays) up to an order of magnitude farther \citep{melosh89}. For a crater of 360~m radius like the one we expect to see on Scheila, the continuous blanket could extend up to 1~km from the center of the crater and patchy ejecta would brighten the surface for a couple of kilometers around the impact feature. The bright spot is thus certainly larger than the continuous blanket but could be explained by the material that fell back on the surface farther away. Estimating the thickness distribution of the ejecta blanket is complex, mainly because very little is known about the physical properties of the regolith on asteroids and its behavior in microgravity. 

\cite{vincent12} have studied the ejecta blanket of one of the most recent impact craters on Asteroid (21) Lutetia, and verified that the semi-empirical cratering laws derived by \citet{housen83} from the Earth and the Moon also applied to the micro-gravity environments on small bodies. One particularly useful empirical relation describes the thickness $B$ of an ejecta blanket as a function of the distance $r$ from the crater center and is written as a power law of the type:
\begin{equation}\label{eq:B_ref1}
\frac{B(r)}{R_{crater}} = K \times \left(\frac{Y}{\rho g R_{crater}}\right)^{e_v/2} \times \left(\frac{r}{R_{crater}}\right)^{-e_r}
\end{equation}
and
\begin{equation}\label{eq:B_ref2}
K = K_4 \times \frac{(e_r-2)}{2\pi}\textrm{sin}(2\theta)^{e_r-2}
\end{equation}
where $e_v = 2e_r - 4$, $K_4$ and $e_r$ are two constants determined for a given combination of target and projectile material, $Y, \rho, g$ are respectively the rock strength, the surface density, and the gravitational acceleration of the asteroid. $\theta$ is the ejection angle, typically equal to 45 degrees.

Typically $e_r$ lies between 2.5 and 3, while $K_4$ is in the range 0.05--0.08. For Lutetia, \cite{vincent12} obtained $e_r = 2.68$ and $K_4 = 0.08$.
The strength of the surface $Y$ was estimated by \cite{Ishiguro:2011to} to be relatively low, equal to 0.3 MPa. The density of the surface $\rho$ is unknown but we can take the standard values of $2500~\textrm{kg}\cdot\textrm{m}^{-3}$ as a first approximation. For this density we calculate a gravity $g = 0.04~\textrm{m}\cdot\textrm{s}^{-2}$.

If we inject these values into Eqs.~(\ref{eq:B_ref1}) and (\ref{eq:B_ref2}) we obtain the following ejecta blanket thickness for Scheila:
\begin{equation}\label{eq:B_scheila}
B(r) = 11 \times \left(\frac{r}{R_{crater}}\right)^{-2.68} \textrm{(meters)}
\end{equation}
where $B(r)$ is the thickness of ejecta at a distance $r$ from the crater center. This relation is shown in Fig.~\ref{Fig4}.

We test this relation against three observations: there must be enough ejecta at a large distance from the crater center to affect the light curve, the velocity needed to reach this distance must be lower than the escape velocity, and the integrated volume of the blanket should be compatible with the amount estimated from the plume analysis.

Since we are studying a microgravity environment we should verify that the material is actually able to reach this maximum distance. We calculated the trajectory of a body ejected by assuming an initial velocity and angle on an airless body. From this ballistic analysis, we derive a velocity $v_0 = 14$ to $24~\textrm{m}\cdot\textrm{s}^{-1}$ for a grain ejected with an angle of 45 degrees to reach a distance of 3.5 to 10 km. The escape velocity, $v_e = \sqrt{2GM_{Scheila}/R_{Scheila}} = 68$~m$\cdot$s$^{-1}$ is nearly an order of magnitude larger and therefore ejecta from the crater can reach the hypothesized limit of the blanket without escaping the asteroid.

Finally, we can integrate Eq.~(\ref{eq:B_ref1}) over the whole ejecta region. Depending on the size of the ejecta region and the crater diameter, we obtain a total volume of ejecta in the range $2.3$ to $5.2 \times 10^6$~m$^3$. From the plume analysis, we had estimated a total volume of ejecta of $1.2$ to $24 \times 10^6$~m$^3$, with 90\% to 99\% of it falling back on the surface. The two numbers are consistent, given the uncertainties we have. The upper limit on the total volume of ejecta comes from the hypothesis that only 1\% of the material actually escaped Scheila and was detected in the plume, but this percentage could be slightly higher. We also have an uncertainty on the albedo/size of the bright patch which constrains our integration limits, and because the law we use for the blanket thickness is usually defined far from the crater center, we tend to overestimate the amount of ejecta deposited close to the rim. Additionally, we cannot know at which distance from the crater the blanket stops being continuous, which can also lead to an overestimation of the deposited volume. However both ranges are in good agreement and it is safe to conclude that a few $10^6$~m$^3$ of material were deposited on the surface.

All results presented in this subsection are summarized in Table \ref{tab:summary}. Based on change in the lightcurve we estimated that the affected surface area was between 3.5 -- 10 km in radius. This is significantly larger than the extent of the continuous ejecta blanket, which we expect to be of order of a kilometer based on the estimated crater size even when using only upper limits. Using Eq.~(\ref{eq:B_scheila}) we calculate a thickness of 2~cm at 3.5~km, and 2~mm at 10~km. This of course raises another question: how thick should it be to have a noticeable effect on the optical properties of the surface? While this is a thin layer, it is larger than the surface scattering length for optical light ($\lesssim 1$~$\mu$m).

\subsection{Source of surface brightening}\label{source}


\noindent What caused the brightening of the surface, and what is the range of albedo changes seen on other bodies in our solar system? First, the increase in albedo could be either endogenic or exogenic, as is nicely illustrated by the recent results from the Dawn mission's investigation of (4) Vesta. Dawn observed localized albedos that ranged from 0.5 to 2 times the asteroid's background albedo \citep{li12,McCord2012}. These spots were mostly found around craters. \citet{McCord2012} suggest that the bright spots are associated with freshly exposed mafic material or impact melt, while the dark spots might be associated carbon-rich deposit from primitive impactors. Another interesting analogue is found in the Martian moons, Phobos and Deimos, which are often classified as D-type asteroids. Topographical studies by Mariner-9, Viking and more recently, HiRISE and Mars Pathfinder showed that bright material is related to craters on both moons  \citep{Thomas1996, Thomas2011}. On Phobos, these features were mostly confined to crater walls, but on Deimos they were visible as kilometer-sized streamers around impact craters \citep{Thomas1996}. Although these features are all clearly associated with impact craters, it is not clear what causes them to be brighter than the background surface material. 

Laboratory experiments indicate that decreasing the particle size of carbonaceous chondrite meteorite samples generally increases albedos by about 30--50\% \citep{JF1973}. Impact melt could add bright material to Scheila's surface. Existing literature suggests that these processes result in at most a factor of 2 increase of the albedo, which would imply a spot radius of $\approx$10~km.

Alternatively, the bright spot might reflect differences within Scheila's regolith - an idea we consider unlikely given that we expect a crater of no more than 150~m deep and that primitive asteroids are generally thought to be undifferentiated. Instead, bright material could be delivered by the impactor. From the plume morphology it was deduced that the impact occurred under a shallow angle \citep{2011ApJ...733L...3B,Ishiguro:2011to}. At the average relative velocity of 5.3 km/s at Scheila's position in the Main Belt, impact models indicate most impactor material will indeed remain solid and that most of it will be deposited outside the crater \citep{Pierazzo2000, Bland2008}. Often colliding bodies have a similar albedo but the increase in brightness can be much larger if the impactor composition is very different from the target. This is particularly relevant to Scheila which is one of the darkest asteroids in the main belt, with an albedo of only 0.0379 \citep{tedesco04}. The highest albedos of asteroids are around 0.4; the average albedo of asteroids at this heliocentric distance (a = 2.9 AU) is $\approx$0.1 \citep{masiero11}, about 3 times larger than Scheila's albedo. This range implies a patch radius between 3.5~--~10~km. Recall that the plume observations indicated a crater radius of 130 -- 360 m, which would typically be surrounded by a continuous ejecta blanket of less than 1 km. Scheila's scar therefore seems very large compared to the scale of the impact. We conclude that the affected area seems large compared to the crater, but that a thin layer (2 mm~--~2cm), possibly mixed with bright impactor material could explain the apparent discrepancy between the photometric measurements and the plume observations.

\begin{table}[!t]%
\begin{center}
\begin{tabular}{l l}
  \hline\hline
	Impactor diameter & 10~--~70 m\\
	Crater diameter & 260~--~720 m\\
	Crater depth & 60~--~150 m\\
	Total mass excavated & $0.3 - 6 \times 10^9$ kg \\
	Diameter bright region & 7 -- 20 km\\
	Minimum thickness of ejecta & 0.2 -- 2 cm\\
  \hline
\end{tabular}
\caption{Summary of the results.}
\label{tab:summary}
\end{center}
\end{table}

\section{Summary and Conclusions}
\noindent The early discovery of (596) Scheila's brightening provided a unique opportunity to directly study a sub-crititical impact on a primitive asteroid. The observed plumes and the measured lightcurve are two independent measures of the scale of the impact and in this paper we tried to connect these two lines of evidence. From the plume observations, Scheila was hit by a 30 -- 80-m sized asteroid creating a sub-km crater. Photometric measurements showed that about a third of Scheila's lightcurve was clearly altered by the impact, which considering the size difference between impactor and target can only be interpreted as a change in its surface properties. In conjunction with plausible albedo increases, the difference between the pre- and post impact lightcurve suggests that a large area of 3.5 -- 10 km in radius was affected by the impact. Cratering physics based on the impactor size suggests that the size of the affected area is larger than expected. Similar albedo changes have been observed on Vesta and the Martian moons, but are not understood. However, most of those observations apply to localized changes, whereas our results suggest that a much larger surface area was affected. Empirical laws describing ejecta blankets however indicate that this large area would be covered by a thin layer of 2 mm to 2 cm thick. This dusting may be mixed with bright impactor material and would be sufficient to explain to observed brightness increase.  

The impact on (596) Scheila demonstrates that a relatively small impact can have significant implications for the optical properties of the surface of a primitive asteroid.  Sub-catastrophic collisions thus need to be taken into account in surface renewal modes. This work should also be taken as a word of caution to the modelers doing shape reconstruction from lightcurve analysis: a small impact can significantly affect the lightcurve of an asteroid, without significantly changing the shape of the body. 

There are currently three space missions in preparation that will target primitive asteroids, e.g., OSIRIS-REx \citep{2012Lauretta}, Hayabusa-2 \citep{2012Abe}, and Marco Polo-R \citep{2009Barucci}. Although they will visit near-Earth asteroids much smaller than Scheila, these missions will shed much more light on the surface processes of primitive asteroids. 
\\
\\
\noindent{\bf Acknowledgements:}
We are thankful to Brian Warner, Masateru Ishiguro, Hiromi Hamanowa, and Marek Hus\'{a}rik for making their photometric measurements available to us. We are grateful to Jessica Sunshine (Univ. Maryland) and the anonymous reviewers for very helpful discussions.

\section*{References}

\bibliography{references,thisissue} \bibliographystyle{icarus}

\begin{thebibliography}{}

\bibitem[{Abe} {\em et~al.}(2012){Abe}, {Yoshikawa}, {Sugita}, {Namiki},
  {Kitazato}, {Okada}, {Tachibana}, {Arakawa}, {Honda}, {Ohtake}, {Tanaka},
  {Fukuhara}, {Takagi}, {Kadono}, {Okazaki}, {Yano}, {Demura}, {Hirata},
  {Nakamura}, {Sawada}, {Mizuno}, {Iwata}, {Saiki}, {Nakazawa}, {Iijima},
  {Hayakawa}, {Kobayashi}, {Mitani}, {Shirai}, {Ogawa}, and {Hayabusa-2 Science
  Team}]{2012Abe}
{Abe}, M., M.~{Yoshikawa}, S.~{Sugita}, N.~{Namiki}, K.~{Kitazato}, T.~{Okada},
  S.~{Tachibana}, M.~{Arakawa}, R.~{Honda}, M.~{Ohtake}, S.~{Tanaka},
  T.~{Fukuhara}, Y.~{Takagi}, T.~{Kadono}, R.~{Okazaki}, H.~{Yano},
  H.~{Demura}, N.~{Hirata}, R.~{Nakamura}, H.~{Sawada}, T.~{Mizuno},
  T.~{Iwata}, T.~{Saiki}, S.~{Nakazawa}, Y.~{Iijima}, M.~{Hayakawa},
  N.~{Kobayashi}, T.~{Mitani}, K.~{Shirai}, K.~{Ogawa},\ and {Hayabusa-2
  Science Team} 2012.
\newblock {Hayabusa-2, C-Type Asteroid Sample Return Mission, Science Targets
  and Instruments}.
\newblock {\em LPI Contributions\/}~{\bf 1667}, 6137.

\bibitem[{Barucci} {\em et~al.}(2009){Barucci}, {Yoshikawa}, {Michel},
  {Kawagushi}, {Yano}, {Brucato}, {Franchi}, {Dotto}, {Fulchignoni}, and
  {Ulamec}]{2009Barucci}
{Barucci}, M.~A., M.~{Yoshikawa}, P.~{Michel}, J.~{Kawagushi}, H.~{Yano}, J.~R.
  {Brucato}, I.~A. {Franchi}, E.~{Dotto}, M.~{Fulchignoni},\ and S.~{Ulamec}
  2009.
\newblock {MARCO POLO: near earth object sample return mission}.
\newblock {\em Experimental Astronomy\/}~{\bf 23}, 785--808.

\bibitem[Bland {\em et~al.}(2008)Bland, Artemieva, Collins, Bottke, Bussey, and
  Joy]{Bland2008}
Bland, P.~A., N.~A. Artemieva, G.~S. Collins, W.~F. Bottke, D.~B.~J. Bussey,\
  and K.~H. Joy 2008.
\newblock {Asteroids on the Moon: Projectile Survival During Low Velocity
  Impact}.
\newblock {\em 39th Lunar and Planetary Science Conference\/}~{\bf 39}, 2045.

\bibitem[Bodewits {\em et~al.}(2011)Bodewits, Kelley, Li, Landsman, Besse, and
  A'Hearn]{2011ApJ...733L...3B}
Bodewits, D., M.~S. Kelley, J.-Y. Li, W.~B. Landsman, S.~Besse,\ and M.~F.
  A'Hearn 2011.
\newblock {Collisional Excavation of Asteroid (596) Scheila}.
\newblock {\em Astrophysical Journal Letters\/}~{\em 733\/}(1), L3.

\bibitem[DeMeo {\em et~al.}(2009)DeMeo, Binzel, Slivan, and
  Bus]{2009Icar..202..160D}
DeMeo, F., R.~P. Binzel, S.~M. Slivan,\ and S.~J. Bus 2009.
\newblock {An extension of the Bus asteroid taxonomy into the near-infrared}.
\newblock {\em Icarus\/}~{\em 202\/}(1), 160--180.

\bibitem[Holsapple {\em et~al.}(2002)Holsapple, Giblin, Housen, Nakamura, and
  Ryan]{2002aste.conf..443H}
Holsapple, K., I.~Giblin, K.~Housen, A.~Nakamura,\ and E.~L. Ryan 2002.
\newblock In {\em {Asteroid Impacts: Laboratory Experiments and Scaling Laws}},
  pp.\  443--462. Asteroids III.

\bibitem[Housen and Holsapple(2012)Housen and Holsapple]{Housen2012}
Housen, K.~R.,\ and K.~A. Holsapple 2012.
\newblock {Craters without ejecta}.
\newblock {\em Icarus\/}~{\em 219\/}(1), 297--306.

\bibitem[{Housen} {\em et~al.}(1983){Housen}, {Schmidt}, and
  {Holsapple}]{housen83}
{Housen}, K.~R., R.~M. {Schmidt},\ and K.~A. {Holsapple} 1983.
\newblock {Crater ejecta scaling laws - Fundamental forms based on dimensional
  analysis}.
\newblock {\em J. Geophys. Res.\/}~{\bf 88}, 2485--2499.

\bibitem[Howell {\em et~al.}(2011)Howell, Lovell, and Jehin]{Howell:2011p2648}
Howell, E., A.~J. Lovell,\ and E.~Jehin 2011.
\newblock {596 Scheila}.
\newblock {\em CBET\/}~(2632), 1--2.

\bibitem[Hsieh {\em et~al.}(2012)Hsieh, Yang, and
  Haghighipour]{2012ApJ...744....9H}
Hsieh, H.~H., B.~Yang,\ and N.~Haghighipour 2012.
\newblock {Optical and Dynamical Characterization of Comet-like Main-belt
  Asteroid (596) Scheila}.
\newblock {\em Astrophysical Journal\/}~{\em 744\/}(1), 9.

\bibitem[Hus{\'a}rik(2012)Hus{\'a}rik]{Husarik:2012wc}
Hus{\'a}rik, M. 2012.
\newblock {Relative photometry of the possible main-belt comet (596) Scheila
  after an outburst}.
\newblock {\em Contributions of the Astronomical Observatory Skalnate
  Pleso\/}~{\bf 42}, 15--21.

\bibitem[Ishiguro {\em et~al.}(2011)Ishiguro, Hanayama, Hasegawa, Sarugaku,
  Watanabe, Fujiwara, Terada, Hsieh, Vaubaillon, and Kawai]{Ishiguro:2011to}
Ishiguro, M., H.~Hanayama, S.~Hasegawa, Y.~Sarugaku, J.~Watanabe, H.~Fujiwara,
  H.~Terada, H.~H. Hsieh, J.~Vaubaillon,\ and N.~Kawai 2011.
\newblock {Observational Evidence for an Impact on the Main-belt Asteroid (596)
  Scheila}.
\newblock {\em Astrophysical Journal Letters\/}~{\bf 740}, L11.

\bibitem[Ishiguro {\em et~al.}(2011)Ishiguro, Hanayama, Hasegawa, Sarugaku,
  Watanabe, Fujiwara, Terada, Hsieh, Vaubaillion, Kawai, Yanagisawa, Kuroda,
  Miyaji, Fukushima, Ohta, Hamanowa, Kim, Pyo, and Nakamura]{Ishiguro:2011uh}
Ishiguro, M., H.~Hanayama, S.~Hasegawa, Y.~Sarugaku, J.-i. Watanabe,
  H.~Fujiwara, H.~Terada, H.~H. Hsieh, J.~J. Vaubaillion, N.~Kawai,
  K.~Yanagisawa, D.~Kuroda, T.~Miyaji, H.~Fukushima, K.~Ohta, H.~Hamanowa,
  J.~Kim, J.~Pyo,\ and A.~M. Nakamura 2011.
\newblock {Interpretation of (596) Scheila's Triple Dust Tails}.
\newblock {\em Astrophysical Journal Letters\/}~{\em 741\/}(1), L24.

\bibitem[Jewitt {\em et~al.}(2011)Jewitt, Weaver, Mutchler, Larson, and
  Agarwal]{Jewitt:2011p2737}
Jewitt, D., H.~Weaver, M.~J. Mutchler, S.~Larson,\ and J.~Agarwal 2011.
\newblock {Hubble Space Telescope Observations of Main-belt Comet (596)
  Scheila}.
\newblock {\em Astrophysical Journal Letters\/}~{\bf 733}, L4.

\bibitem[Johnson and Fanale(1973)Johnson and Fanale]{JF1973}
Johnson, T.~V.,\ and F.~P. Fanale 1973.
\newblock {Optical Properties of Carbonaceous Chondrites and Their Relationship
  to Asteroids}.
\newblock {\em Journal of Geophysical Research\/}~{\bf 78}, 8507--8518.

\bibitem[Larson(2010)Larson]{2010IAUC.9188....1L}
Larson, S.~M. 2010.
\newblock {(596) Scheila}.
\newblock {\em IAU Circular\/}~{\bf 9188}, 1.

\bibitem[{Lauretta} {\em et~al.}(2012){Lauretta}, {Barucci}, {Bierhaus},
  {Brucato}, {Campins}, {Christensen}, {Clark}, {Connolly}, {Dotto}, {Dworkin},
  {Emery}, {Garvin}, {Hildebrand}, {Libourel}, {Marshall}, {Michel}, {Nolan},
  {Nuth}, {Rizk}, {Sandford}, {Scheeres}, and {Vellinga}]{2012Lauretta}
{Lauretta}, D.~S., M.~A. {Barucci}, E.~B. {Bierhaus}, J.~R. {Brucato},
  H.~{Campins}, P.~R. {Christensen}, B.~C. {Clark}, H.~C. {Connolly},
  E.~{Dotto}, J.~P. {Dworkin}, J.~{Emery}, J.~B. {Garvin}, A.~R. {Hildebrand},
  G.~{Libourel}, J.~R. {Marshall}, P.~{Michel}, M.~C. {Nolan}, J.~A. {Nuth},
  B.~{Rizk}, S.~A. {Sandford}, D.~J. {Scheeres},\ and J.~M. {Vellinga} 2012.
\newblock {The OSIRIS-REx Mission - Sample Acquisition Strategy and Evidence
  for the Nature of Regolith on Asteroid (101955) 1999 RQ36}.
\newblock {\em LPI Contributions\/}~{\bf 1667}, 6291.

\bibitem[Lazzarin {\em et~al.}(2006)Lazzarin, Marchi, Moroz, Brunetto, Magrin,
  Paolicchi, and Strazzulla]{2006ApJ...647L.179L}
Lazzarin, M., S.~Marchi, L.~V. Moroz, R.~Brunetto, S.~Magrin, P.~Paolicchi,\
  and G.~Strazzulla 2006.
\newblock {Space Weathering in the Main Asteroid Belt: The Big Picture}.
\newblock {\em Astrophysical Journal\/}~{\em 647\/}(2), L179--L182.

\bibitem[{Li} {\em et~al.}(2012){Li}, {Mittlefehldt}, {Pieters}, {De Sanctis},
  {Schr{\"o}der}, {Hiesinger}, {Russell}, {Raymond}, and {Keller}]{li12}
{Li}, J.-Y., D.~W. {Mittlefehldt}, C.~M. {Pieters}, M.~C. {De Sanctis}, S.~E.
  {Schr{\"o}der}, H.~{Hiesinger}, C.~T. {Russell}, C.~A. {Raymond},\ and U.~H.
  {Keller} 2012.
\newblock {Investigating the Origin of Bright Materials on Vesta: Synthesis,
  Conclusions, and Implications}.
\newblock {\em ACM conference 2012\/}~{\bf 1667}, 6386.

\bibitem[Licandro {\em et~al.}(2011)Licandro, de~L{\'e}on, Kelley, Emery,
  Rivkin, Pinilla-Alonso, Moth{\'e}-Diniz, Campins, and
  Al{\'\i}-Lagoa]{2011epsc.conf.1109L}
Licandro, J., J.~de~L{\'e}on, M.~S. Kelley, J.~Emery, A.~S. Rivkin,
  N.~Pinilla-Alonso, T.~Moth{\'e}-Diniz, H.~Campins,\ and V.~Al{\'\i}-Lagoa
  2011.
\newblock {Multi-wavelength study of activated asteroid (596) Scheila.}
\newblock In {\em EPSC-DPS Joint Meeting 2011}, pp.\  1109. Instituto de
  Astrof{\'\i}sica de Canarias, c/V{\'\i}a L{\'a}ctea s/n, 38200 La Laguna,
  Tenerife, Spain.

\bibitem[{Masiero} {\em et~al.}(2011){Masiero}, {Mainzer}, {Grav}, {Bauer},
  {Cutri}, {Dailey}, {Eisenhardt}, {McMillan}, {Spahr}, {Skrutskie}, {Tholen},
  {Walker}, {Wright}, {DeBaun}, {Elsbury}, {Gautier}, {Gomillion}, and
  {Wilkins}]{masiero11}
{Masiero}, J.~R., A.~K. {Mainzer}, T.~{Grav}, J.~M. {Bauer}, R.~M. {Cutri},
  J.~{Dailey}, P.~R.~M. {Eisenhardt}, R.~S. {McMillan}, T.~B. {Spahr}, M.~F.
  {Skrutskie}, D.~{Tholen}, R.~G. {Walker}, E.~L. {Wright}, E.~{DeBaun},
  D.~{Elsbury}, T.~{Gautier}, IV, S.~{Gomillion},\ and A.~{Wilkins} 2011.
\newblock {Main Belt Asteroids with WISE/NEOWISE. I. Preliminary Albedos and
  Diameters}.
\newblock {\em Astrophysical Journal\/}~{\bf 741}, 68.

\bibitem[McCord {\em et~al.}(2012)McCord, Li, Combe, McSween, Jaumann, Reddy,
  Tosi, Williams, Blewett, Turrini, Palomba, Pieters, De~Sanctis, Ammannito,
  Capria, Le~Corre, Longobardo, Nathues, Mittlefehldt, Schr{\"o}der, Hiesinger,
  Beck, Capaccioni, Carsenty, Keller, Denevi, Sunshine, Raymond, and
  Russell]{McCord2012}
McCord, T.~B., J.-Y. Li, J.~P. Combe, H.~Y. McSween, R.~Jaumann, V.~Reddy,
  F.~Tosi, D.~A. Williams, D.~T.~T. Blewett, D.~Turrini, E.~Palomba, C.~M.
  Pieters, M.~C. De~Sanctis, E.~Ammannito, M.~T. Capria, L.~Le~Corre,
  A.~Longobardo, A.~Nathues, D.~W. Mittlefehldt, S.~E. Schr{\"o}der,
  H.~Hiesinger, A.~W. Beck, F.~Capaccioni, U.~Carsenty, H.~U. Keller, B.~W.
  Denevi, J.~M. Sunshine, C.~A. Raymond,\ and C.~T. Russell 2012.
\newblock {Dark material on Vesta from the infall of carbonaceous volatile-rich
  material}.
\newblock {\em Nature\/}~{\em 490\/}(7422), 83--86.

\bibitem[{Melosh}(1989){Melosh}]{melosh89}
{Melosh}, H.~J. 1989.
\newblock {\em {Impact cratering: A geologic process}}.
\newblock Oxford University Press.

\bibitem[Moreno {\em et~al.}(2011)Moreno, Licandro, Ortiz, Lara,
  Al{\'\i}-Lagoa, Vaduvescu, Morales, Molina, and Lin]{2011ApJ...738..130M}
Moreno, F., J.~Licandro, J.~L. Ortiz, L.-M. Lara, V.~Al{\'\i}-Lagoa,
  O.~Vaduvescu, N.~Morales, A.~Molina,\ and Z.~Y. Lin 2011.
\newblock {(596) Scheila in Outburst: A Probable Collision Event in the Main
  Asteroid Belt}.
\newblock {\em Astrophysical Journal\/}~{\em 738\/}(2), 130.

\bibitem[Nesvorny {\em et~al.}(2005)Nesvorny, Jedicke, Whiteley, and
  Ivezic]{2005Nesvorny}
Nesvorny, D., R.~Jedicke, R.~J. Whiteley,\ and Z.~Ivezic 2005.
\newblock {Evidence for asteroid space weathering from the Sloan Digital Sky
  Survey}.
\newblock {\em Icarus\/}~{\em 173\/}(1), 132--152.

\bibitem[Pierazzo and Melosh(2000)Pierazzo and Melosh]{Pierazzo2000}
Pierazzo, E.,\ and H.~J. Melosh 2000.
\newblock {Understanding Oblique Impacts from Experiments, Observations, and
  Modeling}.
\newblock {\em Annual Review of Earth and Planetary Sciences\/}~{\bf 28},
  141--167.

\bibitem[{Tedesco} {\em et~al.}(2004){Tedesco}, {Noah}, {Noah}, and
  {Price}]{tedesco04}
{Tedesco}, E.~F., P.~V. {Noah}, M.~{Noah},\ and S.~D. {Price} 2004.
\newblock {IRAS Minor Planet Survey V6.0}.
\newblock {\em NASA Planetary Data System\/}~{\bf 12}.

\bibitem[Thomas {\em et~al.}(2011)Thomas, Stelter, Ivanov, Bridges, Herkenhoff,
  and McEwen]{Thomas2011}
Thomas, N., R.~Stelter, A.~Ivanov, N.~T. Bridges, K.~E. Herkenhoff,\ and A.~S.
  McEwen 2011.
\newblock {Spectral heterogeneity on Phobos and Deimos: HiRISE observations and
  comparisons to Mars Pathfinder results}.
\newblock {\em Planetary and Space Science\/}~{\em 59\/}(13), 1281--1292.

\bibitem[Thomas {\em et~al.}(1996)Thomas, Adinolfi, Helfenstein, Simonelli, and
  Veverka]{Thomas1996}
Thomas, P.~C., D.~Adinolfi, P.~Helfenstein, D.~Simonelli,\ and J.~Veverka 1996.
\newblock {The Surface of Deimos: Contribution of Materials and Processes to
  Its Unique Appearance}.
\newblock {\em Icarus\/}~{\em 123\/}(2), 536--556.

\bibitem[{Thomas} {\em et~al.}(2007){Thomas}, {Armstrong}, {Asmar}, {Burns},
  {Denk}, {Giese}, {Helfenstein}, {Iess}, {Johnson}, {McEwen}, {Nicolaisen},
  {Porco}, {Rappaport}, {Richardson}, {Somenzi}, {Tortora}, {Turtle}, and
  {Veverka}]{Thomas2007}
{Thomas}, P.~C., J.~W. {Armstrong}, S.~W. {Asmar}, J.~A. {Burns}, T.~{Denk},
  B.~{Giese}, P.~{Helfenstein}, L.~{Iess}, T.~V. {Johnson}, A.~{McEwen},
  L.~{Nicolaisen}, C.~{Porco}, N.~{Rappaport}, J.~{Richardson}, L.~{Somenzi},
  P.~{Tortora}, E.~P. {Turtle},\ and J.~{Veverka} 2007.
\newblock {Hyperion's sponge-like appearance}.
\newblock {\em Nature\/}~{\bf 448}, 50--56.

\bibitem[Veverka {\em et~al.}(1999)Veverka, Thomas, Harch, Clark, Bell,
  Carcich, Joseph, Murchie, Izenberg, Chapman, Merline, Malin, McFadden, and
  Robinson]{Veverka1999}
Veverka, J., P.~C. Thomas, A.~Harch, B.~Clark, J.~F. Bell, B.~Carcich,
  J.~Joseph, S.~Murchie, N.~Izenberg, C.~Chapman, W.~Merline, M.~Malin,
  L.~McFadden,\ and M.~S. Robinson 1999.
\newblock {NEAR Encounter with Asteroid 253 Mathilde: Overview}.
\newblock {\em Icarus\/}~{\em 140\/}(1), 3--16.

\bibitem[{Vincent} {\em et~al.}(2012){Vincent}, {Besse}, {Marchi}, {Sierks},
  {Massironi}, and {OSIRIS Team}]{vincent12}
{Vincent}, J.-B., S.~{Besse}, S.~{Marchi}, H.~{Sierks}, M.~{Massironi},\ and
  {OSIRIS Team} 2012.
\newblock {Physical properties of craters on asteroid (21) Lutetia}.
\newblock {\em Plan. Space Sci.\/}~{\bf 66}, 79--86.

\bibitem[Warner(2006)Warner]{Warner:vz}
Warner, B.~D. 2006.
\newblock {Scheila Lightcurve}.
\newblock {\em Minor Planet Bulletin\/}~{\bf 33}, 58.

\bibitem[{Warner}(2011){Warner}]{2011MPBu...38...96W}
{Warner}, B.~D. 2011.
\newblock {Upon Further Review: VI. An Examination of Previous Lightcurve
  Analysis from the Palmer Divide Observatory}.
\newblock {\em Minor Planet Bulletin\/}~{\bf 38}, 96--101.

\bibitem[Yang and Hsieh(2011)Yang and Hsieh]{2011ApJ...737L..39Y}
Yang, B.,\ and H.~H. Hsieh 2011.
\newblock {Near-infrared Observations of Comet-like Asteroid (596) Scheila}.
\newblock {\em Astrophysical Journal Letters\/}~{\em 737\/}(2), L39.

\end{thebibliography}


\newpage
{\bf Highlights:}
\begin{itemize}
\item Plume imaging and lightcurves provide independent measures of the impact
\item The impact altered a larger than expected part of the surface
\item Small impacts may play an important role in surface processing of primitive asteroids
 \end{itemize}

\end{document}